\documentclass[conference]{IEEEtran}
\IEEEoverridecommandlockouts
\usepackage[T1]{fontenc}
\usepackage{tikz}
\usetikzlibrary{shapes.geometric, arrows.meta, positioning, calc, backgrounds, fit}
\usepackage{cite}
\usepackage{amsmath,amssymb,amsfonts}
\usepackage{algorithmic}
\usepackage{graphicx}
\usepackage{textcomp}
\usepackage{xcolor}
\usepackage{hyperref} 
\usepackage{orcidlink}
\usepackage{braket}
\usepackage{dsfont}
\usepackage{subcaption}
\usepackage{cite}
\usepackage{amsmath,amssymb,amsfonts}
\usepackage{algorithm}
\usepackage{algorithmic}
\usepackage{textcomp}
\usepackage{mathtools,amsthm}
\usepackage{tcolorbox}
\usepackage{subcaption}
\usepackage{booktabs}
\usepackage{multirow}
\newcommand{\pmark}{\ensuremath{\blacklozenge}}

\newtcolorbox{practitionernote}{
  colback=teal!5,
  colframe=teal!40!black,
  boxrule=0.6pt,
  arc=2pt,
  left=6pt,
  right=6pt,
  top=4pt,
  bottom=4pt
}
\tcbuselibrary{theorems}
\newtcbtheorem[]{mydef}{Definition}%
{colback=blue!5,colframe=blue!35!white,fonttitle=\bfseries}{def}

\def\BibTeX{{\rm B\kern-.05em{\sc i\kern-.025em b}\kern-.08em
    T\kern-.1667em\lower.7ex\hbox{E}\kern-.125emX}}
       \newtheorem{thm}{Theorem}
\def\BibTeX{{\rm B\kern-.05em{\sc i\kern-.025em b}\kern-.08em
    T\kern-.1667em\lower.7ex\hbox{E}\kern-.125emX}}
    
\def\BibTeX{{\rm B\kern-.05em{\sc i\kern-.025em b}\kern-.08em
    T\kern-.1667em\lower.7ex\hbox{E}\kern-.125emX}}
\begin{document}
\bstctlcite{MyControl}
%
\title{Experimental Workflows for Combinatorial Optimization: Towards Quantum Advantage\thanks{We thank Faisal Abu-Khzam for providing his implementation for the LP-reduction rule. This research was supported in part by a National Sciences and Engineering Research Council (NSERC) of Canada Collaborative Research and Training Experience (CREATE) grant on Quantum Computing, an NSERC Alliance Consortium Grant entitled Quantum Software Consortium---Exploring Distributed Quantum Solutions for Canada (QSC), an NSERC Alliance grant on Quantum Computing for Optimal Mobility, and postdoctoral funding provided through University of Victoria. Access to  the \texttt{ibm\_quebec} device used for the hardware executions was made available by Plateforme d' Innovation Numérique et Quantique ($\mathrm{PINQ}^2$). 
\\Corresponding author: pangara@uvic.ca}}

\author{
\orcidlink{0000-0002-9660-011X}Prashanti Priya Angara,$^{1}$
\orcidlink{0000-0003-2045-3456}Luis F.~Rivera,$^{1}$
\orcidlink{0000-0001-9466-7196}Ulrike Stege,$^{1}$
\orcidlink{0000-0002-1004-5830}Hausi M\"uller,$^{1}$
\orcidlink{0000-0002-0617-7195}Ibrahim Shehzad,$^{2}$ 
\orcidlink{0000-0002-2745-1149}Sean Wagner$^{3}$ \\
\small $^{1}$\textit{Department of Computer Science}, \textit{University of Victoria}, Victoria, Canada \\
 \small $^{2}$\textit{IBM Quantum}, \textit{T.~J. Watson Research Center}, Yorktown Heights, New York 10598, USA\\
 \small$^{3}$\textit{IBM Canada}, Markham, ON L6G 1C7, Canada
}

\maketitle

\begin{abstract}

Demonstrating quantum advantage for combinatorial optimization requires more than standalone algorithmic results; it calls for end-to-end case studies that integrate problem modelling, quantum execution, and classical refinement into practical workflows.
This paper presents a sandbox platform for experimenting with hybrid quantum-classical workflows in graph optimization, enabling the systematic study of end-to-end optimization pipelines. Using our platform, we investigate three classically intractable and mutually reducible graph problems—Minimum Vertex Cover, Maximum Independent Set, and Maximum Clique—by transforming them into an unconstrained problem and solving the resulting instances with QAOA on IBM platforms. Our workflow combines classical pre-processing to reduce instance size, quantum optimization on the reduced problem, and classical postprocessing to map quantum outputs to high-quality feasible solutions, thereby avoiding direct constraint encoding in the quantum circuit. We evaluate the approach on synthetic graphs, benchmark instances, and real-world networks, and report hardware experiments on IBM Quantum System One at PINQ² in Bromont, Quebec, powered by IBM’s 156-qubit Heron r2 processor on graphs up to 128 vertices, with circuits involving up to 128 qubits and 13,555 two-qubit gates. 
The results illustrate how sandbox-style end-to-end experimentation can expose bottlenecks, clarify the role of classical-quantum workload partitioning, and provide domain experts and practitioners with a practical guide for interpreting quantum optimization outputs and assessing quantum utility on the road to quantum advantage in combinatorial optimization.
\end{abstract}
 
\vspace*{-5pt}
\section{Introduction}
\label{sec:introduction}
\vspace*{-5pt}

Combinatorial optimization lies at the core of applications spanning logistics, supply chain management, networked infrastructures, manufacturing, energy systems, molecular and materials design, bioinformatics, and broader scientific and engineering discovery~\cite{Bouamama2021, Bai2020, Liang2023, Cardei2005, Nacher2016, SocialCliques2020, Anurag, Abu}. Many such problems are considered classically intractable at scale~\cite{garey1979computers}, motivating the search for more powerful computational approaches. Quantum computing has long promised new computational capabilities for addressing hard optimization problems, most notably through quantum annealing (QA)~\cite{johnson2011quantum} and the hybrid Quantum Approximate Optimization Algorithm (QAOA)~\cite{farhi2014quantum}.

\textit{A.~Toward Demonstrating End-to-End Quantum Advantage.} 
Demonstrating practical quantum advantage remains difficult, especially in realistic end-to-end settings where problem modelling, workflow orchestration, hardware constraints, and classical post-processing must all work together. In many optimization settings, the goal is not just to identify a single best solution, but to obtain several high-quality candidate solutions, for example, in multi-objective decision-making contexts~\cite{multi1,multi2}. In such cases, quantum computing offers a distinctive opportunity because repeated circuit executions naturally produce distributions over candidate solutions rather than a single deterministic output. Our work is motivated by the view that quantum advantage will not emerge from standalone algorithmic results alone, but from experimental platforms that support the systematic study of tightly integrated hybrid workflows. In this paper, we therefore present a \textit{hybrid quantum-classical end-to-end pipeline} realized in a \textit{sandbox platform.}

\textit{B.~Scope and Contributions of this Case Study.} 
Taken together, these elements define an end-to-end, instance-aware hybrid quantum-classical case study for constrained graph optimization. Our contribution is not only a quantum algorithmic experiment, but a broader experimental workflow that integrates classical pre-processing, quantum optimization, and classical post-processing within a modular sandbox environment. This framing enables benchmarking of alternative design choices, studying parameterizations across abstraction layers, and examining how workload should be partitioned between classical and quantum resources in practical settings. This paper presents a domain-specific, end-to-end case study that guides the practitioner to understand and assess quantum utility under realistic computational conditions.

Our main contributions are as follows:
\begin{itemize}
\item \textbf{Instance-aware hybrid classical-quantum pipeline }for constrained combinatorial optimization, designed to produce high-quality solutions and to illustrate how such workflows can serve as practical end-to-end solutions.
\item \textbf{Comprehensive empirical evaluation} across three classes of instances: synthetic graphs, which enable controlled analysis of circuit depth and solution quality; QOBLIB benchmark instances ~\cite{koch2025quantum}, which represent large and classically resistant problem instances; and real-world graphs from biology and collaboration networks, which demonstrate practical applicability.
\item \textbf{Evaluation of the full pipeline on \texttt{ibm\_quebec} quantum hardware}, representing a large-scale demonstration of hybrid quantum-classical optimization for these graph problems and highlighting the value of an effective experimental platform for end-to-end studies.
\item \textbf{Practical guide for domain experts/practitioners} on interpreting quantum optimization pipeline outputs. Unlike classical solvers that typically return a single solution, quantum approaches yield probability distributions over many candidate solutions. We discuss key performance indicators—including distributions over optimal and near-optimal solutions, approximation ratios, and expectation values—and explain their practical meaning in context of constrained combinatorial optimization, offering actionable guidance for practitioners and domain experts.
\end{itemize}

In the remainder of this paper, Section~\ref{sec:framework} presents our case study framework. Section~\ref{sec:background} reviews related work and introduces relevant background and terminology. Section~\ref{sec:methodology} describes our sandbox platform and the graph datasets used in the analysis. Section~\ref{sec:results-1} reports experiments on synthetic graphs  (up to 100 vertices) using the 156-qubit \texttt{ibm\_quebec}  Heron r2 processor. Section~\ref{sec:results-2} presents results on graph instances of up to 4,941 vertices from the Quantum Optimization Benchmarking Library~\cite{koch2025quantum} and Network Repository~\cite{rossi2015network}, evaluated through our hybrid classical-quantum end-to-end pipeline.
\vspace*{-2pt}
\section{Hybrid Case Study Framework}\label{sec:framework}
\vspace*{-2pt}

\textit{A.~Hybrid Sandboxes as Experimental Platforms.} 
Our sandbox platform is organized as a modular three-stage environment comprising a \textit{Classical Pre-processing Sandbox}, a \textit{QAOA Unconstrained Problem Solver Sandbox}, and a \textit{Classical Post-processing Sandbox}. 
Each sandbox encapsulates an exchangeable set of algorithmic components, such as a set of reduction rules used to shrink a problem instance, a circuit ansatz used in quantum execution, and a heuristic used in post-processing. Each can be added, removed, or substituted without changing the behaviour of the other sandboxes. 
This modularity enables controlled benchmarking: a stage can be varied with others are held fixed, allowing observed performance differences to be attributed directly to the modified sandbox component.

\textit{B.~Structural and Parametric Hybridization.} 
The central idea behind our end-to-end pipeline is that classical and quantum processors should each perform the tasks for which they are best suited. In this sense, the pipeline is hybrid in two complementary ways. First, it is \textit{structurally hybrid}: the quantum solver is embedded between classical algorithms that reduce the problem before quantum execution and construct high-quality feasible solutions afterward. Second, it is \textit{parametrically hybrid}: the quantum circuit parameters within QAOA are optimized at runtime by a classical optimizer, as is standard in variational quantum algorithms. Together, these two forms of hybridization enable the workflow to exploit quantum computation where it may provide value, while delegating tractable subproblems to classical methods.

\textit{C.~Benchmarking End-to-End Workflows Across Layers.} 
Grounded in this sandbox platform, we present experimental studies of hybrid quantum-classical computing for constrained combinatorial optimization. The platform allows us to benchmark complete workflows, from problem formulation and encoding to quantum execution and solution analysis. Such experimental sandboxes are particularly important because they support systematic exploration across multiple levels of abstraction, including formulation choices, encoding strategies, algorithmic configurations, orchestration decisions, and refinement steps. They also make it possible to study different blends of classical and quantum methods in a controlled manner, revealing how classical optimization, quantum processing, and iterative feedback can be combined into practical end-to-end pipelines. Using a broad collection of graph-based instances, we leverage our sandbox platform to expose performance bottlenecks, compare hybrid design choices, and identify the conditions under which quantum resources begin to provide meaningful value, utility, and potential advantage. This growing body of experimental evidence illustrates both the promise of quantum methods for combinatorial optimization and the importance of robust platforms that support repeatable benchmarking, cross-sandbox integration, and realistic evaluation. By moving beyond standalone demonstrations to complete computational workflows, this work clarifies what is needed to assess quantum advantage in realistic end-to-end settings.

\textit{D.~Instance-Aware Strategy for Hard Optimization Problems.} 
Our study focuses on three related NP-hard constrained combinatorial graph problems. Although NP-hardness establishes worst-case intractability, it does not imply that every instance is equally resistant to efficient solutions. In practice, small instances or those with favourable structure can often be solved exactly by efficient classical methods, whereas large or structurally unfavourable instances may remain resistant and therefore offer the real opportunity to discover quantum advantage. Real-world instances are rarely uniformly hard; instead, they often contain a mixture of tractable and resistant substructures. This observation motivates a principled, instance-aware strategy that selectively combines classical and quantum methods rather than indiscriminately. A practical consequence is that one should first exhaust what classical methods can solve quickly and exactly, thereby shrinking the instance before offloading the remaining difficulty to quantum hardware. The final step is an informed classical post-processing phase that maps quantum outputs to high-quality feasible solutions. In this view, quantum methods do not replace classical reasoning but rather complement it in targeted ways where they are most likely to add value.

\textit{E.~Graph Problems as Case Studies.} 
As end-user-minded case study for our sandbox platform, we focus on the hard problems \textsc{Minimum Vertex Cover (MinVC)}, \textsc{Maximum Independent Set (MaxIS)}, and \textsc{Maximum Clique (MaxCl)}. These graph problems arise in a wide range of practical settings, including collaboration networks, biological networks, wireless networks, and resource allocation. In wireless networks, for instance, vertices may represent routers, and edges interference relationships; an \textit{independent set} then corresponds to a subset of devices that can operate simultaneously without mutual interference. In resource allocation, vertices may represent resources, and edges indicate conflicts; thus, an independent set represents a conflict-free allocation. A \textit{vertex cover}, in turn, identifies a subset of vertices whose removal leaves pairwise conflict-free resources. In collaboration networks, it may identify key individuals without whom no collaboration edge remains. \textit{Cliques} (vertex subsets that identify complete subgraphs) capture tightly connected groups, such as strongly collaborating teams or dense interaction structures. Since \textsc{MinVC}, \textsc{MaxIS}, and \textsc{MaxCl} are mutually efficiently reducible, they jointly represent a practically relevant family of constrained graph optimization problems.

\textit{F.~The QAOA Constrained Optimization Challenge.} 
QAOA~\cite{farhi2014quantum} is the dominant quantum framework for combinatorial optimization, especially for Unconstrained Binary Optimization (UBO) problems. While this mapping is exact for unconstrained problems, \textsc{MinVC}, \textsc{MaxIS}, and \textsc{MaxCl} involve hard constraints, which are  commonly handled through penalty-based reformulations~\cite{Lucas2014}. Unlike unconstrained problems such as \textsc{MaxCUT}, which map directly and naturally to QUBO formulations, constrained problems require penalty parameters whose values are often problem-specific and highly sensitive, limiting scalability and generalizability~\cite{VERMA2022100594,angara2025scoop}. Moreover, penalty methods impose a binary notion of feasibility: a solution is either feasible or infeasible. This can obscure useful structure in the solution space, since some infeasible solutions may lie close to high-quality feasible ones and can be corrected with only limited post-processing, whereas some feasible solutions may still have poor objective quality. A framework that fails to distinguish between these cases leaves potentially valuable solution information underexploited.

\textit{G.~SCOOP and the \textit{\textsc{MaxPC}} Reformulation.} 
To address this, our platform adopts the SCOOP framework~\cite{angara2025scoop} that guides to reformulate selected constrained combinatorial problems into transformationally equivalent unconstrained variants. Throughout this paper, we consider the unconstrained \textsc{Maximum Profit Cover (MaxPC)} problem~\cite{stege2002}, which is the SCOOP counterpart to \textsc{MinVC}. Because \textsc{MinVC} is mutually reducible to \textsc{MaxIS} and \textsc{MaxCl}, it is a natural representative problem through which to study the broader family of constrained graph problems considered.

\section{Background and Related Work}\label{sec:background}
\vspace{-1pt}
Quantum optimization has seen significant progress toward utility and advantage in recent years~\cite{lanes2025framework, kim2023evidence}. QAOA~\cite{farhi2014quantum} is the canonical gate-based method for combinatorial optimization. Quantum annealing is restricted to quadratic formulations, requiring higher-order terms to be quadratized before embedding~\cite{rosenberg1975reduction}. For constrained combinatorial optimization, the quantum alternating operator ansatz~\cite{hadfield2019quantum} offers an alternative to penalty-based formulations~\cite{Lucas2014} by confining evolution to the feasible subspace via tailored mixing operators that eliminate penalty terms at the cost of increased circuit complexity.
Both approaches for constrained combinatorial optimization carry notable drawbacks: penalty-based formulations risk infeasible solutions, while the quantum alternating operator ansatz trades penalties for circuit complexity. Our prior work, the SCOOP framework~\cite{Angara2025, angara2025scoop}, overcomes both by reformulating selected constrained problems as unconstrained transformational equivalents that are solved with QAOA and polynomial-time classical post-processing---achieving improved approximation ratios and more high quality solutions, and forming the basis for the present work.

For constrained problems related to \textsc{MinVC} and \textsc{MaxIS}, penalty-based and feasibility-preserving approaches have both been explored~\cite{Cook_2020, Bartschi_2020, SaleemTTS23, tomesh2023divide}, as has classical decomposition as a pre-processing step for quantum annealers~\cite{pelofske2019solving, pelofske2023solvingclique}. Notable QAOA variants include QAOA+~\cite{chalupnik2022augmenting}, WS-QAOA~\cite{egger2021warm}, multi-angle QAOA~\cite{herrman2022multi}, and Digitized Counterdiabatic QAOA~\cite{chandarana2022digitized}, each targeting improved approximation ratios or faster convergence. QAOA performance has been studied across Erd\H{o}s-R\'{e}nyi random graphs~\cite{golden2023numerical, SaleemTTS23}, bounded-degree graphs~\cite{shaydulin2023qaoawith, lykov2021performance}, and small structured instances~\cite{herrman2021impact}.

We next introduce the combinatorial optimization problems under study, discuss graph instance reduction, and present the theorem establishing the relationships among the constrained problems and their unconstrained equivalents.

\color{black}

\vspace*{-1.5pt}
\textit{A.~Combinatorial Optimization Problems.} 
A combinatorial optimization problem (COP) $\cal P$ can be formulated as $\min_{\mathbf{x} \in \mathcal{F}} f(\mathbf{x})$, where $\mathcal{F} \subseteq \{0,1\}^n$ is the set of feasible solutions of $\cal P$ and $f(\mathbf{x})$ is the objective function. COP $\cal P$ is \textit{unconstrained} if $\mathcal{F} = \{0,1\}^n$, and \textit{constrained} otherwise. 
An unconstrained COP familiar to many is \textsc{Maximum Cut} \cite{Karp1972}. Due to its native formulation as QUBO \cite{Carlson66}, it is a frequent case study for benchmarking QAOA with simulators and current quantum hardware \cite{farhi2014quantum, Rieffel2018, herrman2021impact}.

\vspace*{-1.5pt}
\textit{B.~Problems of Study.} 
The combinatorial optimization problems we study here are \textsc{Minimum Vertex Cover (MinVC)}, \textsc{Maximum Independent Set (MaxIS)}, and \textsc{Maximum Clique (MaxCl)} \cite{garey1979computers}. 
All three take as input an undirected simple graph $G = (V,E)$. Vertex subset $\textit{VC}\subseteq V$ is a \textit{vertex cover} for $G$ if $\textit{VC}$ \textit{covers} every edge of $G$, i.e.~for edge $uv\in E$ $u\in \textit{VC}$ or $v\in \textit{VC}$. $\textit{IS}\subseteq V$ is an \textit{independent set} for $G$ if no pair of vertices in $\textit{IS}$ shares an edge of $G$, i.e.~for every $u,v\in \textit{IS}$ $uv\notin E$. In contrast, $\textit{C}\subseteq V$ is a \textit{clique} for $G$ if every pair of vertices in $\textit{C}$ shares an edge of $G$, i.e.~for every $u,v\in \textit{C}$ $uv\in E$. Note that the three problems are linear-time equivalent: a subset $V'\subseteq V$ is a vertex cover for $G$ if and only if $V\setminus V'$ is an independent set for $G$ if and only if $V\setminus V'$ is a clique for $\overline{G}$, $G$'s complement graph.\footnote{For an undirected graph $G = (V,E)$, its complement graph $\overline{G}$ is defined as $\overline{G} = (V, (V\times V)\setminus E)$.} The three optimization problems that we set out to solve for a given graph $G$ are \textsc{Minimum Vertex Cover}, where the quest is to determine a smallest vertex cover for $G$, \textsc{Maximum Independent Set}, where we are to determine a largest independent set for $G$, and \textsc{Maximum Clique}, where we look for a largest clique in $G$. 

All three problems are NP-complete~\cite{garey1979computers}. Furthermore, all three problems are constrained COPs: while in each case we look for a subset of vertices, not every subset of $V$ is a vertex cover, an independent set, or a clique for $G$. Because all three problems are linear-time equivalent, and the decision problem \textsc{Vertex Cover} is fixed-parameter tractable when parameterized by its solution size,\footnote{Fastest known fixed-parameter algorithm runs in $O^*\left(1.25284^k\right)$~\cite{improvedVC_STACS24}.}  this work we focus on finding small and minimum vertex covers as our solution strategy; independent sets and cliques can easily be obtained from vertex covers using via some pre- and post-processing.\footnote{\textsc{Independent Set} and \textsc{Clique}, parameterized by solution size, are shown to be W[1]-hard~\cite{downey2013parameterized}.} 

\textit{C.~Reducing Graphs for \textsc{MinVC}.}
\label{SS:RR}
As part of our classical pre- and post-processing steps, we make use of the following polynomial-time reduction rules from parameterized complexity, typically used in pre-processing routines of fixed-parameter algorithms for \textsc{Vertex Cover}, when looking for a vertex cover of size at most $k$~\cite{downey2013parameterized,Cygan2015,Abu}. Here, the minimum vertex cover to be built is named $\text{VC}$. 

    
\textbf{[SR]} Singletons (degree-$0$ vertices) are not included in any minimum vertex cover and therefore can be removed from $G$.

\textbf{[PR]} To cover the (pendant) edge $uw \in E$ incident to a degree-$1$ vertex $u$,  $w$ can be added to $\text{VC}$.

\textbf{[D2R]} If $u\in V$ is a degree-$2$ vertex that is part of a triangle $uvw$ in $G$, i.e.~$uv, vw, wu \in E$, then both $v$ and $w$ can be added to $\text{VC}$. 
A degree-2 vertex $u$ with neighbors $v$ and $w$ that is not part of a triangle can be reduced via \textit{vertex folding}, i.e.~by merging $v$ and $w$ into a single vertex. Note that one has to keep track of this change to later recover the vertex cover for the original input graph.

\textbf{[HDR]} The high-degree reduction rule uses the fact that when looking for a vertex cover of size at most $k>0$, every vertex $u$ of degree larger $k$ must be included into such vertex cover~\cite{Buss1993}.\footnote{Otherwise all of $u$'s neighbors are included into such vertex cover, but then the vertex cover is of size larger than $k$. } When solving \textsc{MinVC} instead of its decision version, we first determine an upper bound for the vertex cover iteratively by creating a greedy method that repeatedly includes a largest degree vertex into the vertex cover until a vertex cover is found. Note that our greedy strategy also evokes [PR] and [D2R] between high degree vertex selections, if applicable. The obtained upper bound is then used to reduce add existing high-degree vertices, if they exist, to $\text{VC}$.  


\textbf{[LPR]} The reduction rule by Linear Programming (LP) relaxes the integer program formulation of \textsc{Vertex Cover} where each vertex $u\in V$ is assigned a value $x_u \in \{0, 1\}$. Here we determine $\min_{\mathbf x}\sum_{u\in V} x_u$  where for every edge $(u, v)\in E$, $x_u + x_v \ge 1$ and $x_u \ge 0$.
Once the LP is solved, $V$ is partitioned into three distinct sets based on their assigned values: $P$ (all $u\in V$ with $x_u > 0.5$), $Q$ (all $u\in V$ with $x_u = 0.5$) and  $R$ (all $u\in V$ with $x_u < 0.5$).
Then $P$ is added to $\text{VC}$, all vertices in $P$ and $R$ are removed from $G$.\footnote{It can be shown that there exists a minimum vertex cover for $G$ that contains every vertex in $P$ and none in $R$.}\\
\indent \textit{D.~Unconstrained SCOOP Problem Twins.} 
\label{SS:SCOOP}
Angara et al.~\cite{angara2025scoop, angara2025flexible} introduce the SCOOP framework and demonstrate how to use it solving a number of constrained COPs: Penalty terms in unconstrained binary formulations are avoided by first deriving a SCOOP problem twin ${\cal P}_U$ of constrained problem ${\cal P}_C$. ${\cal P}_U$ is an unconstrained problem that satisfies a number of properties w.r.t.~${\cal P}_C$ and its unconstrained binary cost function does not require any penalty terms. Solutions to ${\cal P}_U$ imply bounds on existing solutions to ${\cal P}_C$. In particular,  the optimal solution cost for ${\cal P}_U$ lets us directly derive the optimal solution cost for ${\cal P}_C$.   Once solved, solutions to ${\cal P}_U$ can be transformed into solutions to ${\cal P}_C$ by polynomial-time post-processing routines that are guaranteed to not worsen the value of the solution under ${\cal P}_U$'s cost function (improvements are possible, albeit not guaranteed). \\
\indent Each of \textsc{MinVC}, \textsc{MaxIS}, and \textsc{MaxCl} have an unconstrained SCOOP problem twin.  For \textsc{MinVC}, it is \textsc{Maximum Profit Cover (MaxPC)}.\footnote{Unconstrained problem twins of  \textsc{MaxIS}, and \textsc{MaxCl} are \textsc{Maximum Profit Independence}~\cite{vanRooij2003Thesis} and \textsc{Maximum Profit Clique}~\cite{scott2004classical}.}  
Here, any subset $V'\subseteq V$ is a \textit{profit cover}: its profit $\mathfrak{p}_\text{PC}$ for $G$ is evaluated as $\mathfrak{p}_\text{PC}(V') = |E(V')| - |V'|$; $E(V')\subseteq E$ is the subset of edges that are incident to vertices in $V'$. 
For each unconstrained twin, the goal is to determine a subset of vertices that maximizes its respective profit for $G$. All three profit problems are NP-complete, and unconstrained SCOOP twins of their constrained relatives~\cite{angara2025flexible}.\\ 
\indent Since our solution strategy focuses on identifying small vertex covers, we employ \textsc{MaxPC} as the primary unconstrained formulation. The following theorem highlights the relationships between the problems considered here:
\vspace*{-3pt}
  \begin{thm} \cite{garey1979computers,stege2002, scott2004classical, vanRooij2003Thesis}
  \label{theorem1} 
  For any graph $G= (V,E)$, $G$ has a vertex cover $VC \subseteq V$ of size $k$ if and only if $G$ has a subset $\textit{PC} \subseteq V$ with profit $\mathfrak{p}_\text{PC} = |E|-k$ if and only if $G$ has an independent set $\textit{IS} \subseteq V$ of size $|V|-k$ if and only $\overline{G}$ has a clique $\textit{Cl} \subseteq V$ of size $|V|-k$.\\   
 \indent   Furthermore, any subset $V'\subseteq V$ of profit $\mathfrak{p}_\text{PC}$ for $G$ can be post-processed via a classical linear-time algorithm that results in a vertex cover of size at most $|E| - \mathfrak{p}_\text{PC}$.\footnote{In \cite{angara2025flexible}  the equivalences with the problem twins of \textsc{MaxIS} and \textsc{MaxCl} are formalized.} 
    \end{thm}
\vspace*{-3pt}
The choice of solving \textsc{MinVC} and \textsc{MaxPC} is driven by the fixed-parameter tractability of \textsc{Vertex Cover}, which enables effective pre-processing, while solutions to the related  \textsc{MaxIS} and \textsc{MaxCl} can be efficiently obtained through classical polynomial-time post-processing. Eq.~\ref{eq:pcqubo} shows the QUBO for the \textit{\textsc{MaxPC}} problem. Each binary variable $x_v$ has a value  $1$ if $v$ is included in the profit cover $\text{PC}$, and $0$ otherwise.

\begin{tcolorbox}[colback=yellow!5,colframe=yellow!40!black]
\vspace*{-2pt}
    \textbf{Maximize:}
    \begin{align}
    \label{eq:pcqubo}
     \hat{H}_{\text{PC}}(\vec{x}) 
      = \sum_{uv \in E} (x_u + x_v - x_ux_v) - \sum_{v} x_v
      \end{align}
\vspace*{-2pt}
\end{tcolorbox}

\textit{E.~Data.} 
\label{sec:data}
We evaluate our sandbox platform for constrained combinatorial optimization on three classes of graph instances. First, we use synthetic instances generated programmatically~\cite{angara2025flexible} to control graph structure and size, enabling systematic study of algorithm behaviour across varying problem parameters (available at \url{https://github.com/pangara/random-connected-graphs/}). Second, we draw real-world benchmark instances from the Network Repository~\cite{rossi2015network}, introduced by Rossi and Ahmed, that can be solved with the current gate-based qubit limitations. This is a large interactive graph data repository spanning thousands of instances across domains including social, biological, infrastructure, and collaboration networks and covers a wide range of graph sizes and structural properties. Third, we include instances from QOBLIB, the \textit{Quantum Optimization Benchmarking Library (the Intractable Decathlon)} introduced by Koch et al.~\cite{koch2025quantum}, an open-source repository of problem instances and solution records designed for benchmarking quantum optimization algorithms. QOBLIB adopts a model-independent approach, allowing researchers to choose their own formulations, algorithms, and hardware while retaining comparability.

\color{black}


\section{Experimental Setup}
\label{sec:methodology}

\begin{figure*}[!htbp]
    \centering
    \includegraphics[width=0.8\textwidth]{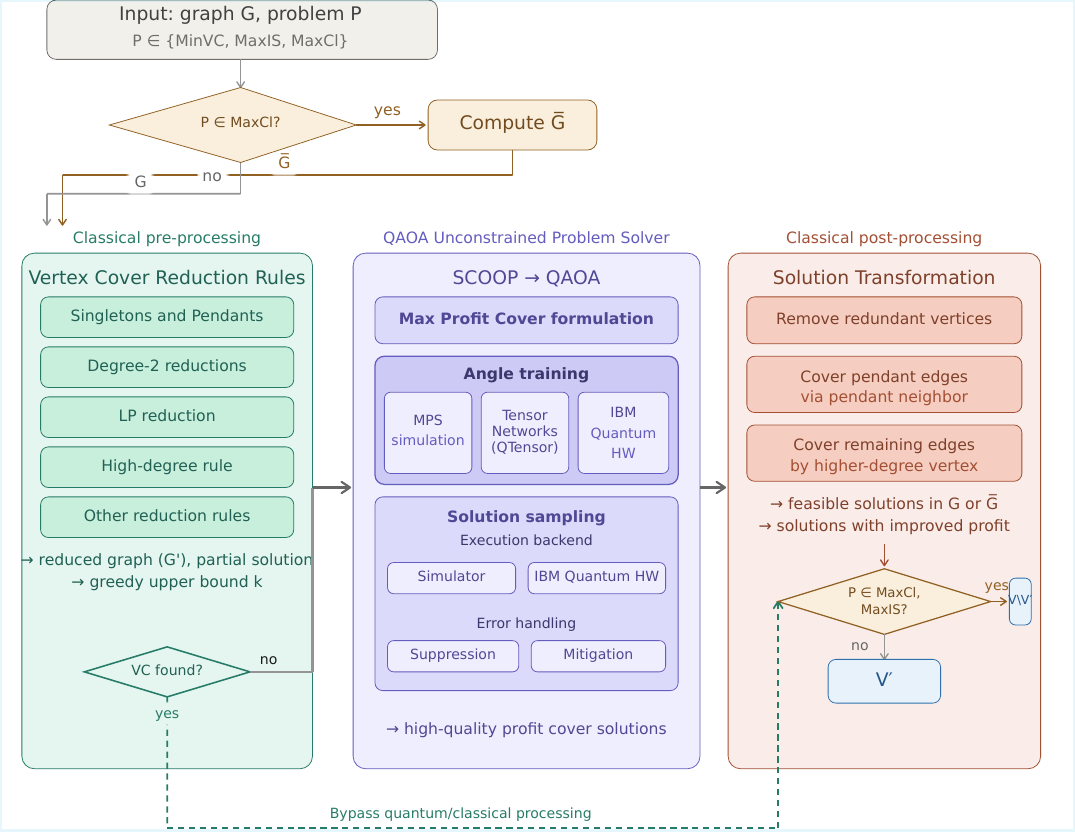}
    \caption{End-to-end classical-quantum pipeline for solving combinatorial optimization problems via the profit framework. The pipeline consists of three stages: (1) classical pre-processing to reduce the graph size, (2) hybrid quantum-classical optimization using the unconstrained QAOA solver on the \textsc{MaxPC} formulation, and (3) classical post-processing to obtain feasible solutions for the original problem. The routing diamonds indicate problem-specific transformations applied at the input and output stages.}
    \label{fig:pipeline}
\end{figure*}

To evaluate determining small vertex covers, large cliques and large independent sets at scale, we implemented the entire sandbox platform (depicted in Figure~\ref{fig:pipeline}). This hybrid classical-quantum end-to-end pipeline consists of three main sandboxes: classical pre-processing via reduction rules (see Section~\ref{SS:RR}), hybrid quantum–classical optimization using the unconstrained QAOA solver using the SCOOP framework~\ref{SS:SCOOP}), and---to transform solutions into feasible solutions for the original problem---classical post-processing using, again, reduction rules (see Section~\ref{SS:RR}) as well as a heuristic step. Namely, when reduction rules do not apply, a greedy step---borrowed from a folklore greedy algorithm---that picks a vertex with maximum many incident uncovered edges and includes it into the vertex cover. Theorem~\ref{theorem1} in Section~\ref{sec:background} establishes the relationship between the target problems---\textsc{MinVC}/\textsc{MaxPC}, \textsc{MaxIS}/\textsc{MaxPI}, and \textsc{MaxCl}/\textsc{MaxPCl}---to \textsc{MaxPC}. This enables us to apply QAOA to the \textsc{MaxPC} formulation, which is natively free of penalty terms, and then obtain feasible solutions for the original problem via post-processing. 

As indicated by the routing diamond at the top of Figure~\ref{fig:pipeline},  \textsc{MaxCl} problems first undergo a complement-graph transformation before entering the main three-stage pipeline; desired solutions are then independent sets for this complement graph. \textsc{MaxVC} and \textsc{MaxIS} problems enter the pipeline directly.

In the following subsections, we describe each stage of the pipeline in detail, along with the graph instances evaluated.

\subsubsection*{Graph instances evaluated} 

 We evaluate our approach on three classes of graph instances, chosen to test scalability across a range of sizes and densities (synthetic graphs), to enable direct comparison with prior quantum and quantum-classical methods (the QOBLIB Decathlon benchmarking library~\cite{koch2025quantum}), and to demonstrate practical applicability on real-world data (application graphs from Network Repository~\cite{rossi2015network}).

\textbf{Synthetic graphs.} For controlled scalability testing, we use random connected Erdős–Rényi graphs~\cite{erdos1960evolution} at edge probabilities of $0.1,0.3,0.5, 0.8$, as well as 3-regular graphs, ranging from 4 to 100 vertices, generated with NetworkX~\cite{hagberg2020networkx}. 

\textbf{QOBLIB Decathlon benchmark.}  We evaluate the \textsc{MaxIS} instances provided by the decathlon benchmark, restricting to graphs of at most 156 vertices to respect the qubit budget of the quantum hardware being used.

\textbf{End-to-end application graphs.} We select graphs from the Network Repository~\cite{rossi2015network} that can be reduced to sizes small enough for near-term quantum hardware yet large enough to require the full pipeline beyond exact classical brute-force methods. This balance enables evaluation in regimes that are both practically meaningful and quantum-feasible.

\textit{A.~Classical Pre-processing Sandbox.} 
\label{subsec:preprocessing}
The goal of pre-processing is to select a vertex set $V_{\text{safe}}$ that we can safely include into a vertex cover without jeopardizing the possibility to obtain an optimal solution. Removing $V_{\text{safe}}$ and the edges incident to it from the graph leaves a reduced graph
 $G'$ whose vertex set is passed to the hybrid quantum-classical solver. Reducing the number of vertices directly reduces the qubit count, which is one of the primary bottlenecks on running large-scale instances on near-term hardware.
 
These rules  are applied in sequence, repeating until no rule fires. While  originally designed to reduce graphs of the decision problem \textsc{Vertex Cover},  in Section \ref{sec:background}~C  we describe how to use them when solving \textsc{MinVC}. Furthermore, all rules can be used when solving its problem twin \textsc{MaxPC}~\cite{stege2002}.  The same rules are applied to all problems---with the problem-specific adaptation that for \textsc{MaxCl} we first will transform the graph instance to its complement-graph.
 
For the end-to-end application graphs considered here, pre-processing alone resolves a large fraction of vertices, yielding smaller graphs that can be run on current quantum hardware. For synthetic graphs, we omit the classical pre-processing step, as it risks fully resolving the instance. The primary objective for the synthetic instances is to assess solver performance consistently across varying graph and layer configurations.

\textit{B.~QAOA Unconstrained Problem Solver Sandbox.} 
\label{sec:scoop-solver}
%
Given a (reduced) graph $G'$, 
we derive its cost Hamiltonian $\hat{H}_{\text{PC}}$ for \textsc{MaxPC}~\cite{angara2025scoop} (Section~\ref{sec:background} {D}) and solve it using QAOA. While the problem is a maximization problem, $\hat{H}_{\text{PC}}$ in Eq.~\ref{eq:pch} is formulated as minimization problem, where vertices in the profit cover evaluated are mapped to $-1$ and the binary variables of vertices not chosen are mapped to $1$. For  $\Delta_{\text{PC}} = \frac{|V|}{2} - \frac{3|E|}{4}$,

\vspace*{-5pt}
\begin{equation}
\label{eq:pch}
\hat{H}_{\text{PC}} = \frac{1}{4}\sum_{uv\in E} (Z_uZ_v + Z_u + Z_v) - \frac{1}{2}\sum_{v\in V}Z_v + \Delta_{\text{PC}}.
\end{equation}

Since the encoding is penalty-term free, no penalty
hyperparameter tuning is required: QAOA operates directly on the unconstrained objective. 
Any solution obtained using for $\hat{H}_{\text{PC}}$  is a profit cover that can (1) either be a vertex cover or (2) can be converted into a solution for any of the target problems by a classical polynomial-time post-processing step, and without loss (but potential gain) of profit (Theorem~\ref{theorem1}).
 
\subsubsection*{\texttt{ibm\_quebec}  Quantum Hardware} 

\textbf{Sequential Training} is done on classical hardware using (1) QTensor, (2) MPS simulator, and (3) \texttt{ibm\_quebec}. For smaller graphs, we run up to $8$  QAOA layers on the MPS simulator. For larger ones we run one layer of QAOA on QTensor tensor network simulator. Each layer is  trained sequentially: first layer 1, then we fix those angles and train layer 2 etc.\footnote{This common approach to training deeper QAOA circuits---used, e.g., to solve Knapsack via QAOA~\cite{Mohseni:2026zwg}---can help mitigate issues with barren plateaus and local minima that arise when optimizing all layers simultaneously~\cite{skolik2021layerwise}.} For \textbf{sampling}, we used the \texttt{SamplerV2} primitive from \texttt{qiskit-ibm-runtime} (v0.43.1), run on \texttt{ibm\_quebec}, 
 accessed via the IBM Quantum cloud platform. Our circuits were built using Qiskit 2.0~\cite{gambetta2022quantum}. 
 Heron r2 arranges qubits in a heavy-hexagonal lattice, where each qubit connects to two or three nearest neighbors; this topology minimizes frequency collisions and spectator errors at cost of reduced connectivity, prioritizing gate fidelity over density. To reduce circuit depth, we transpile our circuits using Qiskit with \texttt{optimization level}  3, together with fractional gate decomposition \cite{fr} enabled. This implements $RZZ(\theta)$ and $RX(\theta)$ natively rather than through multi-gate decompositions. To optimize the device's circuit layout, we run a fresh set of device characterization experiments to get most up-to-date error rates from the backend, right before transpiling.
Our error suppression settings are chosen as follows. Since fractional gates are incompatible with gate twirling in \texttt{qiskit-ibm-runtime}, the available settings are measurement twirling and dynamical decoupling \cite{emdocs}. To select a best combination, we run a mirrored circuit—our circuit composed with its inverse—and measure single-$Z$ observables on all active qubits. We test all combinations (also no error suppression) of the available dynamical decoupling sequences \cite{DD} with measurement twirling both on and off, and then select the combination with mean expectation value across all observables closest to the ideal value of 1.
The 156-qubit capacity of \texttt{ibm\_quebec} directly sets the upper bound on problem size admitted to the quantum solver, motivating the 156-vertex cap applied during instance selection (Section~\ref{sec:data}). 
QAOA circuit parameters $(\boldsymbol{\gamma}, \boldsymbol{\beta})$ are
pre-trained using QTensor~\cite{lykov2021performance}, a tensor network simulator, then fixed
before sampling on \texttt{ibm\_quebec}. \\
%
\indent \textit{C.~Classical Post-processing Sandbox.} 
\label{sec:postprocessing}
%
\vspace{-1pt}
The QAOA output is a bitstring specifying selected vertices in the reduced 
$G'$. Based on the problem, we apply classical post-processing. While this solution is not guaranteed to be feasible for the constrained problem, the following steps, performed classically, yield a valid solution for $G'$,  after which the result is unioned with $V_{\text{safe}}$ (obtained during pre-processing), resulting in a feasible solution for the original constrained problem and input graph $G$. 
In addition to the pool of reduction rules as pre-processing, we inspect the solution obtained for redundant vertices in the cover (a redundant vertex is one whose neighborhood is covered by a subset of the cover and therefore can be removed) as well as for uncovered edges that are not treated by any of the other reduction rules. We cover such edges in a greedy manner: for an uncovered edge $uv$ we pick $u$ if the degree of uncovered edges for $u$ is higher than the one for $v$, and $v$ otherwise.  
%
%
%
%
%
%
%
After post-processing, the solution for $G'$ is combined with $V_{\text{safe}}$ to
obtain a feasible vertex cover for $G$. For \textsc{MaxCl} or \textsc{MaxIS} problems, the output diamond in
Figure~\ref{fig:pipeline} routes the solution through a final inversion ($V \setminus V'$) to translate the cover 
into an independent set for $G$ or a clique for $\bar{G}$. 

\vspace*{-3pt}
\section{Results Guide I: QAOA Unconstrained Quantum Solver Performance}
\begin{figure*}[!htbp]
    \centering
    \begin{subfigure}[b]{0.32\textwidth}
        \centering
        \includegraphics[width=\textwidth]{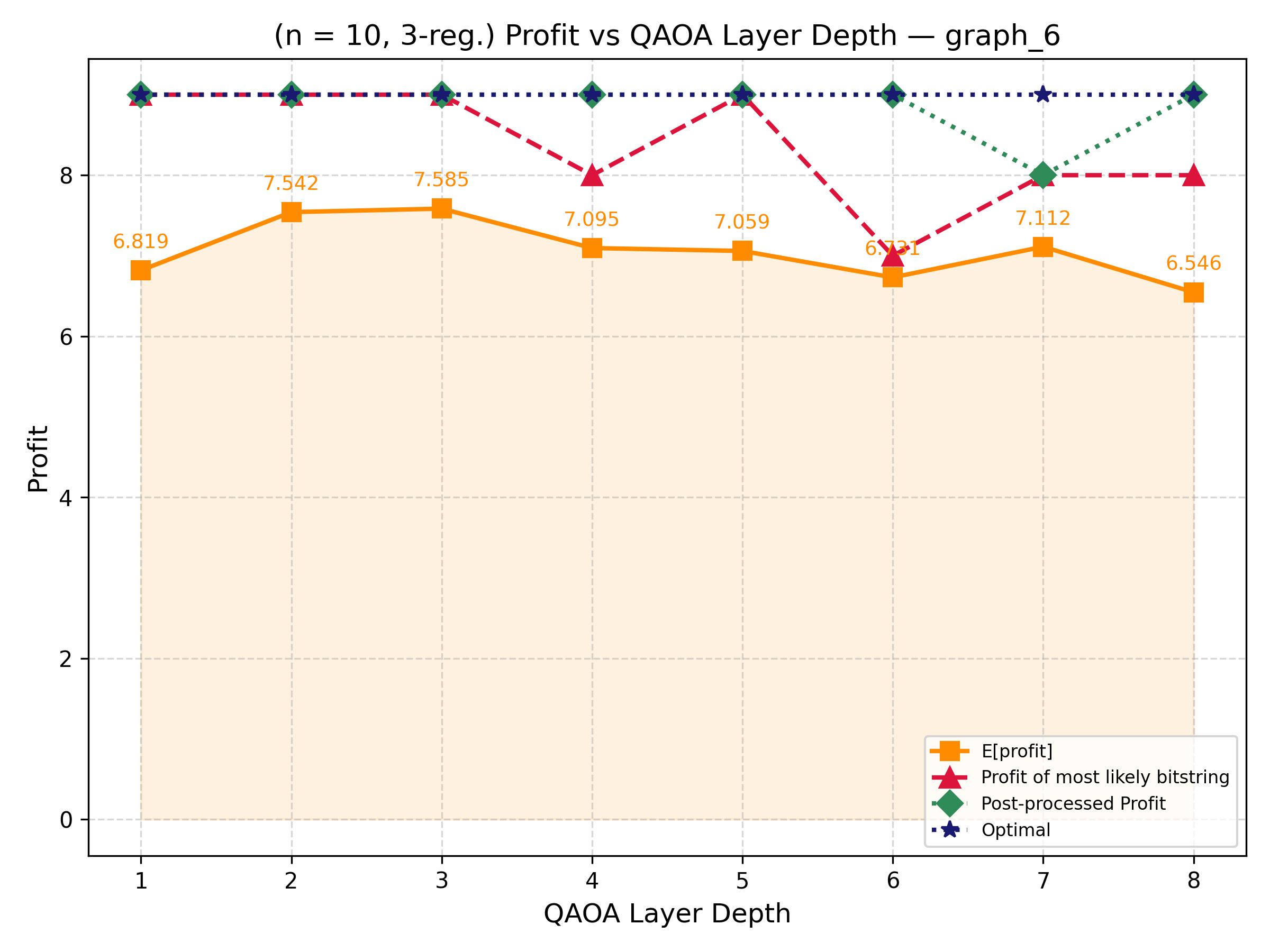}
        \caption{}
        \label{fig:qaoa-single-profit}
    \end{subfigure}
    \hfill
    \begin{subfigure}[b]{0.32\textwidth}
        \centering
        \includegraphics[width=\textwidth]{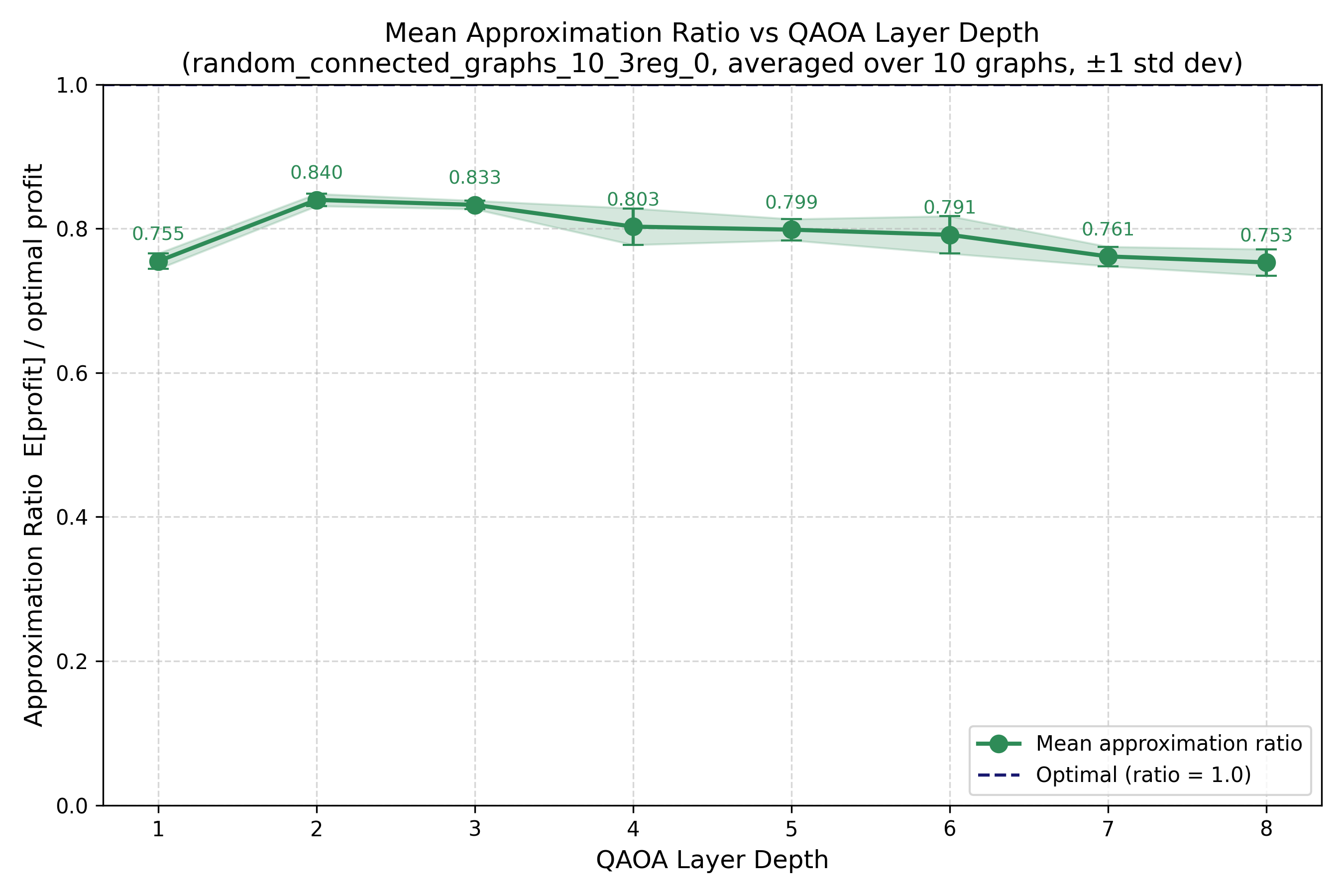}
        \caption{}
        \label{fig:qaoa-approx-ratio}
    \end{subfigure}
    \hfill
    \begin{subfigure}[b]{0.32\textwidth}
        \centering
        \includegraphics[width=\textwidth]{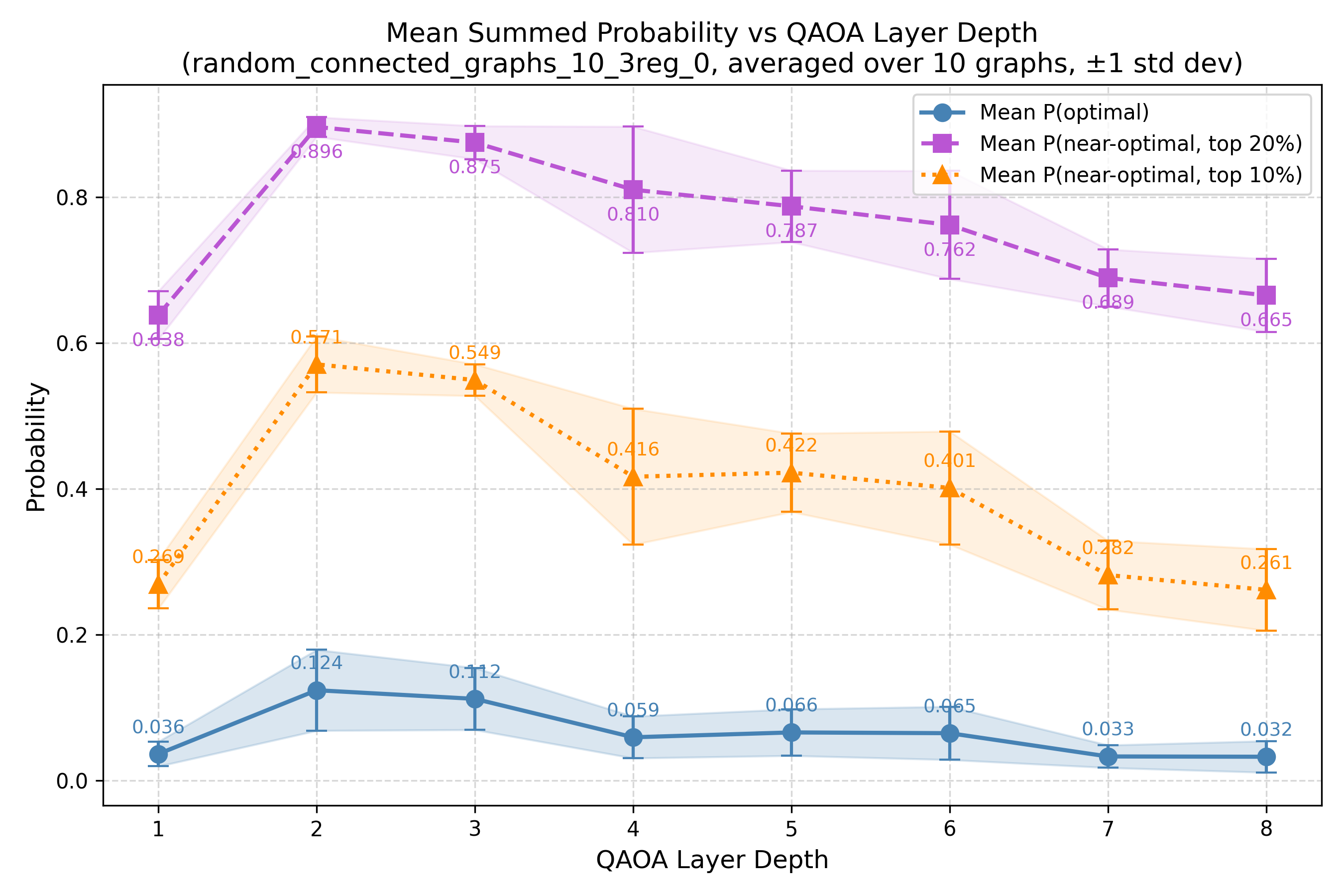}
        \caption{}
        \label{fig:qaoa-aggregate-summed}
    \end{subfigure}
    \caption{QAOA solver performance on synthetic graph instances. \textbf{(a)} \textsc{MaxPC} profit comparison for a representative single graph instance: optimal profit, most likely sampled profit, probability-weighted average profit,  post-processed profit. \textbf{(b)} Approximation ratio for 3-regular graphs, $n=10$. \textbf{(c)} Aggregate summed probability across all synthetic graph instances, with mean \& variance of probability mass on optimal \& near-optimal solutions ($\geq 90\%$-optimal, $\geq 80\%$-optimal) across QAOA $p$ layers.}
    \label{fig:qaoa-results}
\end{figure*}

\vspace*{-2pt}
\label{sec:results-1}
\begin{practitionernote}
\textbf{\pmark} In the following sections, this symbol marks explanations intended for domain experts and practitioners.
\end{practitionernote}
We now evaluate the performance for \textsc{MaxPC} on a set of small synthetic graphs using the unconstrained QAOA solver. 
We conduct experiments using only the QAOA Solver Sandbox, with the Pre-processing Sandbox inactive. We report results with and without post-processing. 
This allows us to assess the performance of QAOA independently of any classical pre- or post-processing. 
We use synthetic graphs spanning a range of structural regimes---from sparse to dense, as well as three-regular graphs---and evaluate the QAOA solver performance for \textsc{MaxPC} directly on these  (both the Pre- and the Post-processing Sandbox are inactive).
We evaluate solver performance both via per-instance analysis and via aggregate metrics---specifically, the approximation ratio and the probability mass assigned to optimal and near-optimal solutions---each capturing a complementary aspect of the output distribution. Figure~\ref{fig:qaoa-results} presents three complementary views of QAOA performance across the synthetic graph instances. 
\begin{practitionernote}
\textbf{\pmark}
Figure~\ref{fig:qaoa-single-profit} summarizes the practitioner’s decision landscape:  \emph{optimal profit} is used as reference solution,  \emph{most likely sampled profit} (mode) reflects the highest-probability bitstring,  \emph{weighted average profit} captures overall distribution quality, and  \emph{post-processed profit} represents final pipeline performance. The expected value provides a direct measure of solution quality; for \textsc{MinVC} on a graph with $|E|$ edges, $|E| - \mathbb{E}[\text{profit}]$ corresponds to the average vertex cover size obtained from the samples. 
\end{practitionernote}
\paragraph{Approximation Ratio}
Figure~\ref{fig:qaoa-approx-ratio} depicts the \emph{approximation ratio}, which measures the quality of the best solution obtained from the QAOA output distribution relative to a known classical optimum, capturing how close the solver gets to the optimal objective value. A key advantage of \textsc{MaxPC} being unconstrained  is that the approximation ratio is a reliable and unambiguous performance indicator.
\begin{practitionernote}
\textbf{\pmark}
Often, combinatorial constraints are enforced by including penalty terms into the objective function, yielding a penalized Hamiltonian. This introduces a subtle distortion: the cost landscape for near-optimal \emph{feasible} solutions may be misaligned with the cost landscape for near-optimal \emph{bitstrings}, since infeasible bitstrings with low penalty may rank higher than high-quality feasible solutions~\cite{angara2025flexible}. Thus, approximation ratios can be  inflated:  constraint-violating solutions may achieve deceptively favorable scores. In contrast, approximation ratios for the unconstrained \textsc{MaxPC} directly reflect the intrinsic quality of the sampled solutions. This highlights an important advantage of unconstrained formulations and motivates independent evaluation of the QAOA Solver Sandbox, decoupled from classical pre- and post-processing.
\end{practitionernote}

\paragraph{Summed optimal and near-optimal probabilities}
This metric aggregates the total probability mass that the QAOA circuit assigns to near-optimal or optimal bitstrings. We compute three cumulative probability curves: the probability mass on the optimal solution(s), the mass on solutions achieving at least $90\%$ of the optimal objective value, and the mass on solutions achieving at least $80\%$ of the optimal objective value. 
Figure~\ref{fig:qaoa-aggregate-summed} aggregates the summed probability curves across all synthetic graph instances in the benchmark. While individual instances exhibit variance due to differences in graph structure (density, regularity, and size), the aggregate trends confirm that our formulation consistently concentrates probability mass on the high-quality solution subspace as circuit depth increases.

\begin{practitionernote}
\textbf{\pmark}
A framework that performs well on a single hand-picked instance but degrades on others offers limited practical assurance. The aggregate curves show that, even as optimal probability remains modest across instances, the $\geq 90\%$-optimal mass is meaningfully elevated across the full benchmark---our pipeline captures good solutions reliably, not just occasionally. Domain practitioners should treat this aggregate view as the primary evidence of pipeline reliability, reserving single-instance plots for diagnostic and interpretive purposes. Variance across instances also provides a practical signal: high variance in summed probability curves across graphs of similar size suggests sensitivity to instance structure.
\end{practitionernote}

A key parameter in QAOA is the number of alternating layers $p$, which controls circuit depth and expressibility. Deeper circuits can in principle better approximate the target ground state, but are subject to increased noise on real hardware and longer simulation times classically. In this work, we evaluate at up to $p = 8$ layers, which allows us to assess convergence behaviour and distribution concentration at greater depth. This is made tractable by the small graph sizes used in this section, which remain within the feasibility limits of simulation. Scaling to larger instances introduces well-known bottlenecks on both classical simulators (exponential state space) and real quantum hardware (decoherence, gate error accumulation)---a point we return to in later sections when discussing the role of the Pre-processing Sandbox in reducing instance size prior to quantum solving. Additionally, our pipeline incorporates more sophisticated classical post-processing than is typical in quantum benchmarking suites such as QOBLIB, allowing us to extract higher-quality solutions from the raw distribution.

Taken together, these results establish that the QAOA Solver Sandbox, operating independently of classical pre- and post-processing, produces output distributions that are meaningfully concentrated on high-quality solutions for the synthetic instances considered here. The exact approximation ratio, while useful as a scalar summary, does not fully characterise solver behaviour: a complete assessment requires examining the shape of the output distribution, the evolution of probability concentration with circuit depth, and the interplay between quantum sampling and classical post-processing. The subsequent results section~\ref{sec:results-2} introduces the full pipeline and demonstrates how each stage---pre-processing, quantum solving, and post-processing---contributes to overall solution quality for larger and more challenging instances.


\section{Results Guide II: End-to-End Performance}
\label{sec:results-2}
\vspace{-1pt}

Real-world data is often modelled as graphs; many of the most critical and frequent problems that apply to real-world graphs are NP-hard or otherwise computationally intractable.  For such problems and their instances, classically obtained solutions---where feasible---typically feature exactly one solution to the problem at hand. 
Quantum algorithms through repeated circuit executions naturally produce distributions over candidate solutions. 
With quantum computing moving towards quantum advantage and having reached quantum utility for selected applications, sandbox platforms as proposed here may support a user in exploring quantum utility for their needs. While many graphs obtained from real-data sets might not be considered hard instances as they can be solved to satisfaction in a tractable time (some even in polynomial time), others require exponential running time when solved classically, making their solvability infeasible at scale. Although quantum hardware still comes with its limitations with respect to qubit count, 2-qubit gate count and circuit depth, instances that can be run on quantum hardware now reach sizes that become interesting when solving harder instances of computationally hard problems.  Even hard instances often can safely be reduced in size---an important goal for the algorithmic engineering task is to decide when and what to offload to the quantum computer. 

In our case study, we investigated a number of graph instances from the Network Repository database~\cite{rossi2015network}. While some of the instances were solvable optimally solely through polynomial-time pre-processing through the reduction rules discussed (Section \ref{SS:RR}), others were first reduced as much as possible before solving the reduced instance on the quantum computer. In the latter case, we followed up through post processing. Results are reported in Table~\ref{table:end-to-end}.
Having established the performance characteristics of the QAOA solver on \textsc{MaxPC} in isolation, we now evaluate the full hybrid pipeline end-to-end, with all three sandboxes active. In contrast to Section~\ref{sec:results-1}, where synthetic instances were passed directly to the quantum solver, the experiments presented here use the complete workflow: graph instances first pass through the Pre-processing Sandbox, where classical reduction techniques shrink the problem prior to quantum execution, before entering the QAOA Unconstrained Problem Solver Sandbox, and finally the Post-processing Sandbox, where classical algorithmic methods or heuristics operate on the quantum output distribution to build high-quality, feasible solutions for the constrained problem. The central question of this section is therefore not how well QAOA performs as a standalone solver, but rather what solution quality the pipeline as a whole can deliver once classical pre- and post-processing are permitted to contribute. 
We organise our evaluation across two  datasets~\cite{koch2025quantum,rossi2015network} in Section~VI-A and Section~VI-B, respectively.

\paragraph{Quantum Circuit statistics: depth and two-qubit gate count}
The quantum circuit implementing QAOA at $p$ layers has a depth and a number of two-qubit (CNOT or CZ) gates that grow with $p$ and with the number of edges in the problem graph. On near-term hardware, these are the primary drivers of noise accumulation, and they are directly reduced by the Pre-processing Sandbox shrinking the graph prior to encoding. We report both for every instance.

\begin{practitionernote}
\textbf{\pmark}~Deeper circuits with more two-qubit gates accumulate more noise and degrade solution quality. Pre-processing mitigates this by reducing the graph, and therefore the circuit, before quantum execution.
\end{practitionernote}
\vspace{-1pt}
\paragraph{Pipeline boundary, when quantum execution is invoked}
Unlike the isolated solver experiments (Section~\ref{sec:results-1}), the full pipeline may resolve a graph through classical pre-processing, invoking the quantum solver only when a non-trivial reduced instance remains. We flag this in our tables.

\textit{A.~QOBLIB Decathlon Benchmark.} 
\label{subsec:qoblib}
 We evaluated our pipeline on all QOBLIB instances for the \textsc{MaxIS} problem with up to 128 nodes; instances beyond this size exceed the 156-qubit capacity of the target hardware after encoding.\footnote{To solve \textsc{MaxIS}, we employ the unconstrained \textsc{MaxPC} formulation and defer conversion to feasible \textsc{MaxIS} solutions to classical post-processing.}

Table~\ref{table:qoblib} summarizes each instance by graph size, achieved objective values ($p=0$ vs.\ $p=1$ QAOA), resulting \textsc{MaxIS} solution quality (best, likely, optimal, and approximation ratio, $\alpha_{\mathrm{best}}$, for $p=1$), circuit complexity (two-qubit depth and gates), runtime (training and sampling), and applied error suppression strategy. Profit gain over random sampling---while modest---is significant in the near-term regime, when circuit depths are large and noise-prone. The fact that $p=1$ QAOA yields consistent improvements under these constraints highlights its practical potential on current hardware. 


\begin{practitionernote}
\textbf{\pmark}~Raw profit denotes the objective value of the highest-probability sampled bitstring, post-processed profit reflects the optimal value obtained by the classical refinement pipeline. This improvement incurs negligible overhead making post-processing an integral component of the overall solution procedure.
\end{practitionernote}

\textit{B.~End-to-End Evaluation on Real-World Graph Instances.} 
\label{subsec:e2e-networkrepository}
%
Our end-to-end results reflect as key nuance of NP-hardness: instance difficulty varies significantly in practice. Consistent with this, in our case study on real-world graphs from~\cite{rossi2015network}, some instances were solved optimally entirely through polynomial-time pre-processing (Section~\ref{SS:RR}~C), others were first reduced as much as possible before solving the reduced graph on the quantum computer. In the latter case, we followed up through post-processing. Results are reported in Table~\ref{table:end-to-end}. This hybrid, instance-aware pipeline directly realizes the observation that real-world problems have a mix of tractable and resistant structure, using classical efficiency where possible and reserving quantum resources where needed.
For each real-world graph (Table~\ref{table:end-to-end}) we compare our pipeline's final solution against two external reference points where available: (i)~the best solution reported in~\cite{rossi2015network}, and (ii)~any additional known results from the literature.

\begin{table*}[!htbp]
\centering
\caption{\textbf{QAOA performance on QOBLIB instances for \textsc{MaxIS}.} `Graph' columns report instance name, and graph size ($|V|$, $|E|$). `Best Profit' shows best objective value  for \textsc{MaxPC} obtained by random sampling ($0$ layers of QAOA) and by QAOA at depth $p=1$.  `\textsc{MaxIS} Solution' reports size of the independent set for layer 1, corresponding to  best sampled solution (Best),  most frequently sampled solution (Likely),  optimal solution size reported in  QOBLIB (Opt),  approximation ratio of best sampled solution ($\alpha_{\text{best}} = \text{Best}/\text{Opt}$). `QAOA Circuit' lists circuit complexity as two-qubit depth, total number of two-qubit gates. `Time (s)' reports classical training time on QTensor ($t_{\text{train}}$), sampling time  on \texttt{ibm\_quebec} ($t_{\text{samp}}$).  `ES' indicates used error suppression (none, measurement-twirling only (\texttt{mt\_only}), or dynamical decoupling sequences such as~\texttt{dd\_XpXm}).}
\label{table:qoblib}

\setlength{\tabcolsep}{3pt}
\renewcommand{\arraystretch}{1.1}

\resizebox{\textwidth}{!}{%
\begin{tabular}{@{}rcc|cc|cccc|cc|cc|c@{}}
\toprule

 \multicolumn{3}{c|}{\textbf{Graph}}
  & \multicolumn{2}{c|}{\textbf{Best Profit}}
  & \multicolumn{4}{c|}{\textbf{\textsc{MaxIS} Solution}}
  & \multicolumn{2}{c|}{\textbf{QAOA Circuit}}
  & \multicolumn{2}{c|}{\textbf{Time (s)}}
  & \multirow{2}{*}{\textbf{ES}} \\

 Name & $|V|$ & $|E|$
  & Rnd & $p=1$
  & Best & Likely & Opt & $\alpha_{\text{best}}$
  & 2Q depth & 2Q gates
  & $t_{\text{train}}$ & $t_{\text{samp}}$
  & \\

\midrule

 C125-9 & 125 & 787 & 632 & 649 & 28 & 27 & 34 & 0.82 & 1469 & 10507 & 5102.6 & 410 & \texttt{mt\_only} \\
 
 aves-sparrow-social & 52 & 454 & 404 & 408 & 13 & 13 & 13 & 1.0 & 808 & 2842 & 2334.8 & 345 & \texttt{none} \\
 
 chesapeake & 39 & 170 & 144 & 145 & 16 & 15 & 17 & 0.94 &  353 & 1076 & 576.3 & 294 & \texttt{none} \\
 es60fst01 & 123 & 159 & 74 & 77 & 56 & 55 & 60 & 0.93 & 168 & 954 & 178.7 & 271 & \texttt{dd\_XpXm} \\
 farm & 17 & 39 & 31 & 32 & 10 & 10 & 10 & 1.0 & 89 & 174 & 72.2 & 268 & \texttt{none} \\
 insecta-ant-colony1-day38 & 56 & 1134 & 1034 & 1024 & 5 & 4 & 6 & 0.83 & 3168 & 8994 & 25033.8 & 606 & \texttt{none} \\
 karate & 34 & 78 & 61 & 62 & 20 & 20 & 20 & 1.0 & 202 & 426 & 148.0 & 279 & \texttt{none} \\
 mammalia-kangaroo-interactions & 17 & 91 & 78 & 78 & 4 & 4 & 4 & 1.0 & 118 & 367 & 199.5 & 270 & \texttt{none} \\
 aves-sparrow-social & 52 & 454 & 404 & 408 & 13 & 13 & 13 & 1.0 & 808 & 2842 & 2334.8 & 345 & \texttt{none} \\
 sloane\_1dc\_128 & 128 & 1471 & 1139 & 1204 & 12 & 12 & 16 & 0.75 & 2556 & 13555 & 8190.8 & 520 & \texttt{mt\_only} \\
 sloane\_1dc\_64 & 64 & 543 & 465 & 471 & 8 & 8 & 10 & 0.8 & 1114 & 3873 & 1704.6 & 372 & \texttt{none} \\
 sloane\_1zc\_128 & 128 & 1120 & 878 & 886 & 17 & 16 & 18 & 0.94 & 1816 & 11659 & 4502.5 & 443 & \texttt{mt\_only} \\

\bottomrule
\end{tabular}%
}

\end{table*}

\begin{table*}[!htbp]
\centering
\caption{End-to-end results of the classical–quantum pipeline. Instances are reduced via pre-processing ($|V_{\text{safe}}|$, $G'$ or $\bar{G}$), solved by QAOA unless \emph{Solved} (\emph{Bypassed}), and refined via post-processing to obtain feasible solutions for $\mathcal{P}$ (\textsc{MaxIS}, \textsc{MaxCl}, \textsc{MinVC}). When $\mathcal{P}=\textsc{MaxCl}$, the pipeline is applied to $\bar{G}$, with $|\bar{E}|$ reported instead of $|E|$. }

\label{table:end-to-end}

\setlength{\tabcolsep}{3pt}
\renewcommand{\arraystretch}{1.1}

\resizebox{\textwidth}{!}{
\begin{tabular}{@{}rccc|cc|c|cccc|c@{}}
\toprule

\multicolumn{4}{c|}{\textbf{Graph}} 
& \multicolumn{2}{c|}{\textbf{Classical Pre-Proc.}} 
& \textbf{QAOA Solver} 
& \multicolumn{4}{c|}{\textbf{Classical Post-Proc.}} 
& \multirow{2}{*}{\textbf{Ref.}} \\

Name & $|V|$ & $|E|$ & $|\bar{E}|$
& $|V_{\text{safe}}|$ & $G'(V, E)$ or $\bar{G}'(V, \bar{E})$ 
& Best $\mathfrak{p}_{PC}$ 
& $\mathfrak{p}_{PC}$ & $\cal P$ & $|\text{Sol}|$ &  $|\text{Sol}|_{\text{known}}$
& \\

\midrule
 aves-sparrow-social & 52 &  & 872
 & 7 & $(45, 566)$
 & 525
 & 530 & $\textsc{MaxCl}$ & 9 & 10
 & \cite{rossi2015network}\\


 bio-celegans & 453 &  & 100353
 & 402 & $(51, 892)$
 & $814$
 & $847$ & $\textsc{MaxCl}$ & $6$ & $9$
 & \cite{rossi2015network}\\

 bio-diseasome & 516 & 1188 &
 & 236 & $(96, 201)$
 & $123$
 & $125$ & $\textsc{MinVC}$ & $312$ & 285 
 & \cite{gao2016fixed} \\

 bio-diseasome & 516 &  & 131682
 & 505 & Solved
 & Bypassed
 & Bypassed & $\textsc{MaxCl}$ & 11 & 11
 & \cite{rossi2015network}\\


 bio-yeast & 1458 &  & 1060205
 & 1452 & Solved
 & Bypassed
 & Bypassed & $\textsc{MaxCl}$ & 6 & 6 
 & \cite{rossi2015network}\\

 ca-netscience & 379 & 914 &
 & 173 & $(86, 195)$
 & $126$
 & $130$ & $\textsc{MinVC}$ & $238$ & 214 
 & \cite{zhu2025optimizing} \\

 ca-netscience & 379 &  & 70717
 & 370 & Solved
 & Bypassed
 & Bypassed & $\textsc{MaxCl}$ & 9 & 9
 & \cite{rossi2015network}\\

 chesapeake & 39 & 170 &
 & 9 & $(25, 49)$
 & 34
 & 36 & $\textsc{MaxIS}$ & 17 & 17
 &  \cite{koch2025quantum}\\

 chesapeake & 39 &  & 571
 & 0 & $(39, 571)$
 & $528$
 & $535$ & $\textsc{MaxCl}$ & 3 & 5
 & \cite{rossi2015network}\\

 es60fst01 & 123 & 159 &
 & 44 & $(34, 46)$
 & 26
 & 27 & $\textsc{MaxIS}$ & 60 & 60
 & \cite{koch2025quantum}\\

 es60fst02 & 186 & 280 &
 & 32 & $(124, 193)$
 & 103
 & 118 & $\textsc{MaxIS}$ & 79 & 88
 & \cite{koch2025quantum}\\

 es60fst03 & 113 & 142 &
 & 18 & $(74, 98)$
 & $26$
 & $54$ & $\textsc{MaxIS}$ & 51 & 55
 & \cite{koch2025quantum}\\

 es60fst04 & 162 & 238 &
 & 39 & $(85, 132)$
 & $73$
 & $81$ & $\textsc{MaxIS}$ & 72 & 78
 & \cite{koch2025quantum}\\

 farm & 17 & 39 &
 & 7 & Solved
 & Bypassed
 & Bypassed & $\textsc{MaxIS}$ & 10 & 10
 & \cite{koch2025quantum}\\

 farm & 17 &  & 97
 & 1 & $(16, 82)$
 & $69$
 & $69$ & $\textsc{MaxCl}$ & 3 & 3
 & \cite{rossi2015network}\\

 football & 35 & 118 &
 & 16 & Solved
 & Bypassed
 & Bypassed & $\textsc{MaxIS}$ & 16 & $16$
 & \cite{koch2025quantum}\\

 football & 35 &  & 477
 & 29 & Solved
 & Bypassed
 & Bypassed & $\textsc{MaxCl}$ & 6 & 6
 & \cite{rossi2015network}\\

 ia-email-univ & 1133 &  & 635827
 & 1121 & Solved
 & Bypassed
 & Bypassed & $\textsc{MaxCl}$ & 12 & 12
 & \cite{rossi2015network}\\

 ia-enron-only & 143 & 623 &
 & 25 & $(94, 371)$
 & $232$
 & $306$ & $\textsc{MinVC}$ & $90$ & 86
 &  \cite{zhu2025optimizing}\\

 ia-enron-only & 143 &  & 9530
 & 87 & $(56, 1223)$
 & $1131$
 & $1175$ & $\textsc{MaxCl}$ & 8 & 8
 & \cite{rossi2015network}\\

 ia-fb-messages & 1266 & 6451 &
 & 548 & $(64, 81)$
 & 40
 & 44 & $\textsc{MinVC}$ & 585 & 578 
 & \cite{zhu2025optimizing}\\

 ia-infect-dublin & 410 &  & 81080
 & 378 & $(32, 145)$
 & $125$
 & $129$ & $\textsc{MaxCl}$ & \textbf{16} & \textbf{9}
 & \cite{rossi2015network}\\

 ia-infect-hyper & 113 & 2196 &
 & 1 & $(111, 2098)$
 & $1862$
 & $2005$ & $\textsc{MinVC}$ & $94$ & 93
 &  \cite{cai2017finding}\\

 inf-power & 4941 &  & 12197676
 & 4929 & $(12, 30)$
 & $24$
 & $24$ & $\textsc{MaxCl}$ & 6 & 6 
 & \cite{rossi2015network}\\

 insecta-ant-colony1-day38 & 56 & 1134 &
 & 2 & $(53, 1048)$
 & $969$
 & $1000$ & $\textsc{MaxIS}$ & 6 & 6
 & \cite{koch2025quantum}\\

 insecta-ant-colony1-day38 & 56 &  & 406
 & 7 & $(47, 210)$
 & 178
 & 185 & $\textsc{MaxCl}$ & \textbf{24} & \textbf{22}
 & \cite{rossi2015network}\\

 karate & 34 & 78 &
 & 13 & $(4, 4)$
 & 2
 & 2 & $\textsc{MaxIS}$ & 19 & 20
 & \cite{koch2025quantum} \\

 karate & 34 &  & 483
 & 12 & $(22, 176)$
 & $156$
 & $158$ & $\textsc{MaxCl}$ & 4 & 5
 & \cite{rossi2015network}\\

 mammalia-kangaroo-interactions & 17 & 91 &
 & 7 & $(8, 17)$
 & $11$
 & $11$ & $\textsc{MaxIS}$ & 4 & 4
 & \cite{koch2025quantum}\\

 mammalia-kangaroo-interactions & 17 &  & 45
 & 7 & Solved
 & Bypassed
 & Bypassed & $\textsc{MaxCl}$ & \textbf{10} & \textbf{9}
 & \cite{rossi2015network}\\

 rt-retweet & 96 & 117 &
 & 32 & Solved
 & Bypassed
 & Bypassed & $\textsc{MinVC}$ & 32 & 32
 &  \\

 rt-retweet & 96 &  & 4443
 & 92 & Solved
 & Bypassed
 & Bypassed & $\textsc{MaxCl}$ & 4 & 4
 & \cite{rossi2015network}\\

\bottomrule
\end{tabular}
}

\end{table*}

The end-to-end evaluation illustrates complementary roles of the three sandboxes. Instances with rich local structure are heavily reduced (Pre-processing Sandbox), sometimes no quantum execution is required at all. For structurally harder residual graphs, the QAOA Solver Sandbox is invoked on the reduced graph; in the case that the obtain solution so far is not feasible for the constrained problem, the Post-processing Sandbox builds a feasible solution to the constrained problem and then stitches partial solutions (from the pre-processing stage) into full-instance solutions. The resulting solutions are competitive with---and in some cases superior to---the best solutions reported in the Network Repository~\cite{rossi2015network}.

\begin{practitionernote}
\textbf{\pmark}~Results reported in~\cite{rossi2015network} represent solutions contributed by the community using mature classical methods (often heuristics), making them a meaningful external baseline. Where our pipeline matches or exceeds these values, it demonstrates practical competitiveness---not merely performance on synthetic benchmarks. 
\end{practitionernote}

\section{Conclusions and Future Work}
\label{sec:conclusions}
\vspace*{-1pt}




This paper presents a hybrid quantum-classical end-to-end pipeline for combinatorial optimization, focusing on graph-based problems. The pipeline, implemented as a set of modular sandboxes, integrates classical pre-processing, a QAOA solver, and classical post-processing, enabling controlled benchmarking and systematic exploration of hybrid workflows.


Our hybrid pipeline represents a principled step \emph{towards} quantum advantage. The pipeline was evaluated on synthetic graphs, QOBLIB benchmark graphs, and real-world instances, using the \texttt{ibm\_quebec} quantum hardware. 
%
Specifically, we find that: (i) classical pre-processing via problem reduction is essential to bring problem instances within reach of current quantum hardware; (ii) QAOA delivers competitive solution quality on constrained instances where exact classical methods begin to struggle; and (iii) classical post-processing plays an indispensable role in the pipeline---by offloading feasibility enforcement, and local refinement back to classical compute---
a productive division of labour.

Several promising directions emerge from this work, including 
swapping in alternative solvers and pre- and post-processors across the three pipeline stages; conducting spectral analysis of graph instances to leverage structural characteristics that govern circuit depth requirements and solution quality 
and generalizing the pipeline to 
problem classes involving higher-order terms. 

Our sandbox platform is a starting point that serves as a practical tool for algorithm designers, domain experts and practitioners, looking to explore and benchmark hybrid approaches on the path towards quantum advantage.

\newpage
\bibliographystyle{IEEEtran}
\bibliography{IEEEabrv, scoop}

@article{multi2,
    author = {Andrew King},
    title = {Multi-objective optimization by quantum annealing},
    journal = {arXiv:2511.01762},
    year = {2025}
}

@article{multi1,
    author = {Ayse Kotil and Elijah Pelofske and Stephanie Riedmüller and Daniel J.~Egger and Stephan Eidenbenz and Thorsten Koch and Stefan Woerner},
    title = {Quantum approximate multi-objective optimization},
    journal = {Nature Computational Science},
    volume = {5},
    pages = {1168--1177},
    year = {2025}
}

@inproceedings{pelofske2019solving,
  title={Solving large minimum vertex cover problems on a quantum annealer},
  author={Pelofske, Elijah and Hahn, Georg and Djidjev, Hristo},
  booktitle={Proceedings of the 16th ACM International Conference on Computing Frontiers},
  pages={76--84},
  note={2019}
}

@article{VERMA2022100594,
title = {Penalty and partitioning techniques to improve performance of {QUBO} solvers},
journal = {Discrete Optimization},
volume = {44},
pages = {100594},
year = {2022},
issn = {1572-5286},
doi = {https://doi.org/10.1016/j.disopt.2020.100594},
url = {https://www.sciencedirect.com/science/article/pii/S1572528620300281},
author = {Amit Verma and Mark Lewis}
}

@article{Anurag,
    author = {Anurag Verma and Austin Buchanan and Sergiy Butenko },
    title = {Solving the Maximum Clique and Vertex Coloring Problems on Very Large Sparse Networks},
    journal = {INFORMS Journal on Computing},
    volume = {27},
    number = {1},
    pages = {164--177},
    year = {2015}
}

@article{SocialCliques2020,
    author = {Fox, Jacob and Roughgarden, Tim and Seshadhri, C. and Wei, Fan and Wein, Nicole},
    title = {Finding Cliques in Social Networks: A New Distribution-Free Model},
    journal = {SIAM Journal on Computing},
    volume = {49},
    number = {2},
    pages = {448-464},
    year = {2020},
    doi = {10.1137/18M1210459},
    URL = {https://doi.org/10.1137/18M1210459},
    eprint = {https://doi.org/10.1137/18M1210459}
}

@INPROCEEDINGS{Abu,
  author={Abu-Khzam, F.N. and Langston, M.A. and Suters, W.H.},
  booktitle={The 3rd ACS/IEEE International Conference onComputer Systems and Applications, 2005.}, 
  title={Fast, effective vertex cover kernelization: a tale of two algorithms}, 
  year={2005},
  volume={},
  number={},
  pages={16-22},
  doi={10.1109/AICCSA.2005.1387015}}

@article{Buss1993,
    author = {Buss, J. F. and Goldsmith, J.},
    title = {Nondeterminism within {$P^*$}},
    journal = {SIAM Journal on Computing},
    volume = {22},
    number = {3},
    pages = {560--572},
    year = {1993}
}

@book{downey2013parameterized,
  title={{Fundamentals of Parameterized Complexity}},
  author={Downey, Rodney G and Fellows, Michael R},
  volume={4},
  note={2013},
  publisher={Springer}
}

@article{Rieffel2018,
    author = {Wang, Z. and Hadfield, S. and Jiang, Z. and Rieffel, E.G.},
    title = {Quantum approximate optimization algorithm for {MaxCut}: A fermionic view},
    journal = {Physical Review A},
    year = {2018}
}

@article{Carlson66,
    author ={Carlson, R. C. and Nemhauser, G. L.},
    title = {Scheduling to Minimize Interaction Cost},
    journal = {Operations Research},
    volume = {14},
    issue = {1},
    pages = {52--58},
    year = {1966}
}

@mthesis{vanRooij2003Thesis,
  title={{Tractable Cognition: Complexity Theory in Cognitive Psychology}},
  author={van Rooij, Iris},
  school = "University of Victoria",
  note= "{PhD} Thesis, 2003"
}

@InProceedings{improvedVC_STACS24,
  author =	{Harris, David G. and Narayanaswamy, N. S.},
  title =	{{A Faster Algorithm for Vertex Cover Parameterized by Solution Size}},
  booktitle =	{41st International Symposium on Theoretical Aspects of Computer Science (STACS 2024)},
  pages =	{40:1--40:18},
  series =	{Leibniz Intern.~Proc.~in Informatics (LIPIcs)},
  ISBN =	{978-3-95977-311-9},
  ISSN =	{1868-8969},
  year =	{2024},
  volume =	{289},
  URN =		{urn:nbn:de:0030-drops-197508},
  doi =		{10.4230/LIPIcs.STACS.2024.40}
  }

@Inbook{Karp1972,
    author="Karp, Richard M.",
    title="Reducibility among Combinatorial Problems",
    bookTitle="Complexity of Computer Computations: Proceedings of a symposium on the Complexity of Computer Computations",
    year="1972",
    publisher="Springer US",
    pages="85--103",
    isbn="978-1-4684-2001-2",
    doi="10.1007/978-1-4684-2001-2_9",
    url="https://doi.org/10.1007/978-1-4684-2001-2_9"
}

@book{Cygan2015,
year = {2015},
author = {Cygan, Marek. and Fomin, Fedor V. and Kowalik, Lukasz. and Lokshtanov, Daniel. and Marx, Dániel. and Pilipczuk, Marcin. and Pilipczuk, Michał. and Saurabh, Saket.},
address = {Cham},
edition = {1st ed. 2015.},
isbn = {3-319-21275-3},
language = {eng},
publisher = {Springer International Publishing},
title = {Parameterized Algorithms},
}

@article{Nacher2016,
author = {Nacher, Jose C. and Akutsu, Tatsuya},
address = {United States},
copyright = {2016 Elsevier Inc.},
issn = {1046-2023},
journal = {Methods},
language = {eng},
pages = {57-63},
publisher = {Elsevier},
title = {Minimum dominating set-based methods for analyzing biological networks},
volume = {102},
year = {2016},
}

@article{Bai2020,
author = {Bai, Xin and Zhao, Danning and Bai, Sen and Wang, Qiang and Li, Weilue and Mu, Dongmei},
copyright = {2019 Elsevier B.V.},
issn = {1570-8705},
journal = {Ad Hoc Networks},
language = {eng},
pages = {102023},
publisher = {Elsevier},
title = {Minimum connected dominating sets in heterogeneous {3D} wireless ad hoc networks},
volume = {97},
year = {2020},
}

@article{Bouamama2021,
  author       = {Salim Bouamama and Christian Blum},
  title        = {An Improved Greedy Heuristic for the Minimum Positive Influence Dominating Set Problem in Social Networks},
  journal      = {Algorithms},
  volume       = {14},
  year         = {2021},
  number       = {3},
  pages        = {79},
  issn         = {1999-4893},
  doi          = {10.3390/a14030079},
}

@article{Liang2023,
	title = {Construction of node- and link-fault-tolerant virtual backbones in wireless networks},
	volume = {79},
	issn = {1573-0484},
	doi = {10.1007/s11227-023-05180-9},
	language = {en},
	number = {12},
	urldate = {2024-08-27},
	journal = {The Journal of Supercomputing},
	author = {Liang, Jiarong and Zeng, Weijian and Du, Xiaojiang},
	month = aug,
	year = {2023},
	pages = {13050--13074},
}

@article{Cardei2005,
author = {Cardei, M and Du, DZ},
address = {DORDRECHT},
copyright = {Springer Science + Business Media, Inc. 2005},
issn = {1022-0038},
journal = {Wireless networks},
language = {eng},
number = {3},
pages = {333-340},
publisher = {Springer Nature},
title = {Improving wireless sensor network lifetime through power aware organization},
volume = {11},
year = {2005},
}

@article{Lucas2014,
author = {Lucas, Andrew},
issn = {2296-424X},
journal = {Frontiers in Physics},
language = {eng},
publisher = {Frontiers Media S.A},
title = {Ising formulations of many {NP} problems},
volume = {2},
year = {2014},
}

@article{farhi2014quantum,
  title={A quantum approximate optimization algorithm},
  author={Farhi, Edward and Goldstone, Jeffrey and Gutmann, Sam},
  journal={arXiv preprint arXiv:1411.4028},
  year={2014}
}

@article{hadfield2019quantum,
  title={From the quantum approximate optimization algorithm to a quantum alternating operator ansatz},
  author={Hadfield, Stuart and Wang, Zhihui and O’Gorman, Bryan and Rieffel, Eleanor G and Venturelli, Davide and Biswas, Rupak},
  journal={Algorithms},
  volume={12},
  number={2},
  pages={34},
  year={2019},
  publisher={MDPI}
}

@article{egger2021warm,
  title={Warm-starting quantum optimization},
  author={Egger, Daniel J and Mare{\v{c}}ek, Jakub and Woerner, Stefan},
  journal={Quantum},
  volume={5},
  pages={479},
  year={2021},
  publisher={Verein zur F{\"o}rderung des Open Access Publizierens in den Quantenwissenschaften}
}

@article{herrman2022multi,
  title={Multi-angle quantum approximate optimization algorithm},
  author={Herrman, Rebekah and Lotshaw, Phillip C and Ostrowski, James and Humble, Travis S and Siopsis, George},
  journal={Scientific Reports},
  volume={12},
  number={1},
  pages={6781},
  year={2022},
  publisher={Nature Publishing Group UK London}
}

@inproceedings{lykov2021performance,
  title={Performance evaluation and acceleration of the {QTensor} quantum circuit simulator on {GPUs}},
  author={Lykov, Danylo and Chen, Angela and Chen, Huaxuan and Keipert, Kristopher and Zhang, Zheng and Gibbs, Tom and Alexeev, Yuri},
  booktitle={2021 IEEE/ACM Second International Workshop on Quantum Computing Software (QCS)},
  pages={27--34},
  note={2021}
}

@article{SaleemTTS23,
  author       = {Saleem,Zain H.  and
                   Tomesh, Teague and
                   Tariq, Bilal and
                   Suchara, Martin},
  title        = {Approaches to Constrained Quantum Approximate Optimization},
  journal      = {{Springer Nature} Computer Science},
  volume       = {4},
  number       = {2},
  pages        = {183},
  year         = {2023},
  doi          = {10.1007/S42979-022-01638-4},
  bibsource    = {dblp computer science bibliography, https://dblp.org}}

@inproceedings{tomesh2023divide,
  title={Divide and conquer for combinatorial optimization and distributed quantum computation},
  author={Tomesh, Teague and Saleem, Zain H and Perlin, Michael A and Gokhale, Pranav and Suchara, Martin and Martonosi, Margaret},
  booktitle={2023 IEEE International Conference on Quantum Computing and Engineering (QCE)},
  volume={1},
  pages={1--12},
  note={2023}
}

@inproceedings{shaydulin2023qaoawith,
  title={{QAOA} with $N\cdot p\geq 200$},
  author={Shaydulin, Ruslan and Pistoia, Marco},
  booktitle={2023 IEEE International Conference on Quantum Computing and Engineering (QCE)},
  volume={1},
  pages={1074--1077},
  note={2023}
}

@inproceedings{golden2023numerical,
  title={Numerical evidence for exponential speed-up of {QAOA} over unstructured search for approximate constrained optimization},
  author={Golden, John and B{\"a}rtschi, Andreas and O'Malley, Daniel and Eidenbenz, Stephan},
  booktitle={2023 IEEE International Conference on Quantum Computing and Engineering (QCE)},
  volume={1},
  pages={496--505},
  note={2023}
}

@article{herrman2021impact,
  title={Impact of graph structures for {QAOA} on {MaxCut}},
  author={Herrman, Rebekah and Treffert, Lorna and Ostrowski, James and Lotshaw, Phillip C and Humble, Travis S and Siopsis, George},
  journal={Quantum Information Processing},
  volume={20},
  number={9},
  pages={289},
  year={2021},
  publisher={Springer}
}

@article{Angara2025,
  title={The Art of Avoiding Constraints: A Penalty-free Approach to Constrained Combinatorial Optimization with {QAOA}},
  author={Angara, Prashanti Priya and Lykov, Danylo and Stege, Ulrike and Alexeev, Yuri and M{\"u}ller, Hausi},
  journal={arXiv preprint arXiv:2503.10077},
  year={2025}
}

@article{johnson2011quantum,
  title={Quantum annealing with manufactured spins},
  author={Johnson, Mark W and Amin, Mohammad HS and Gildert, Suzanne and Lanting, Trevor and Hamze, Firas and Dickson, Neil and Harris, Richard and Berkley, Andrew J and Johansson, Jan and Bunyk, Paul and others},
  journal={Nature},
  volume={473},
  number={7346},
  pages={194--198},
  year={2011},
  publisher={Nature Publishing Group UK London}
}

@InProceedings{stege2002,
author="Stege, Ulrike
and van Rooij, Iris
and Hertel, Alex
and Hertel, Philipp",
editor="Bose, Prosenjit
and Morin, Pat",
title="An ${O}(pn + 1.151^p)$-Algorithm for $p$-Profit Cover and Its Practical Implications for Vertex Cover",
booktitle="Algorithms and Computation",
note="2002",
publisher="Springer",
pages="249--261",
isbn="978-3-540-36136-7"
}

@article{koch2025quantum,
  title={Quantum Optimization Benchmarking Library-The Intractable Decathlon},
  author={Koch, Thorsten and Neira, David E Bernal and Chen, Ying and Cortiana, Giorgio and Egger, Daniel J and Heese, Raoul and Hegade, Narendra N and Cadavid, Alejandro Gomez and Huang, Rhea and Itoko, Toshinari and others},
  journal={arXiv:2504.03832},
  year={2025}
}

@book{garey1979computers,
  title={Computers and Intractability},
  author={Garey, Michael R and Johnson, David S},
  volume={174},
  note={1979},
  publisher={Freeman San Francisco}
}

@misc{scott2004classical,
  title={Classical and parameterized complexity of cliques and games},
  author={Scott, Allan Edward Jolicoeur},
  school = "University of Victoria",
  note= "{M}aster's Thesis",
  year = {2004}
}

@article{lanes2025framework,
  title={A Framework for Quantum Advantage},
  author={Lanes, Olivia and Beji, Mourad and Corcoles, Antonio D and Dalyac, Constantin and Gambetta, Jay M and Henriet, Loic and Javadi-Abhari, Ali and Kandala, Abhinav and Mezzacapo, Antonio and Porter, Christopher and others},
  journal={arXiv preprint arXiv:2506.20658},
  year={2025}
}

@article{kim2023evidence,
  title={Evidence for the utility of quantum computing before fault tolerance},
  author={Kim, Youngseok and Eddins, Andrew and Anand, Sajant and Wei, Ken Xuan and Van Den Berg, Ewout and Rosenblatt, Sami and Nayfeh, Hasan and Wu, Yantao and Zaletel, Michael and Temme, Kristan and others},
  journal={Nature},
  volume={618},
  number={7965},
  pages={500--505},
  year={2023},
  publisher={Nature Publishing Group UK London}
}

@inproceedings{angara2025scoop,
  title={{SCOOP}: A quantum-computing framework for constrained combinatorial optimization},
  author={Angara, Prashanti Priya and Martins, Emily and Stege, Ulrike and M{\"u}ller, Hausi},
  booktitle={2025 IEEE International Conference on Quantum Computing and Engineering (QCE)},
  volume={1},
  pages={65--75},
  year={2025},
  organization={IEEE}
}

@article{rosenberg1975reduction,
  title={REDUCTION OF BIVALENT MAXIMIZATION TO THE QUADRATIC CASE.},
  author={Rosenberg, Ivo G},
  year={1975}
}

@article{hagberg2020networkx,
  title={Networkx: Network analysis with python},
  author={Hagberg, Aric and Conway, Drew},
  journal={URL: https://networkx. github. io},
  volume={1031},
  year={2020}
}

@article{erdos1960evolution,
  title={On the evolution of random graphs},
  author={Erdos, Paul and R{\'e}nyi, Alfr{\'e}d and others},
  journal={Publ. Math. Inst. Hung. Acad. Sci},
  volume={5},
  number={1},
  pages={17--60},
  year={1960},
  publisher={Citeseer}
}

@inproceedings{rossi2015network,
  title={The network data repository with interactive graph analytics and visualization},
  author={Rossi, Ryan and Ahmed, Nesreen},
  booktitle={Proceedings of the AAAI conference on artificial intelligence},
  volume={29},
  number={1},
  year={2015}
}

@article{gambetta2022quantum,
  title={Quantum-centric supercomputing: The next wave of computing},
  author={Gambetta, Jay},
  journal={IBM Research Blog},
  year={2022}
}

@misc{fr,
  title = {{Fractional Gates}},
  howpublished = {\url{https://quantum.cloud.ibm.com/docs/en/guides/fractional-gates}},
  year = {2026},
  note = {Accessed: April 20, 2026}
}

@misc{emdocs,
  title = {{Error mitigation and suppression techniques}},
  howpublished = {\url{https://quantum.cloud.ibm.com/docs/en/guides/error-mitigation-and-suppression-techniques}},
  year = {2026},
  note = {Accessed: April 20, 2026}
}

@misc{DD,
  title = {{Dynamical Decoupling Options}},
  howpublished = {\url{https://quantum.cloud.ibm.com/docs/en/api/qiskit-ibm-runtime/options-dynamical-decoupling-options#sequence_type}},
  year = {2026},
  note = {Accessed: April 20, 2026}
}

@article{Mohseni:2026zwg,
    author = "Mohseni, Naeimeh and Houle, Julien-Pierre and Shehzad, Ibrahim and Cortiana, Giorgio and O'Meara, Corey and Watts, Adam Bene",
    title = "{Constrained Quantum Optimization at Utility Scale: Application to the Knapsack Problem}",
    journal = "arXiv.2603.00260",
    archivePrefix = "arXiv",
    primaryClass = "quant-ph",
    year = "2026"
}

@phdthesis{angara2025flexible,
  title={Flexible integration of classical and quantum techniques in the evolutionary path to quantum utility},
  author={Angara, Prashanti Priya},
  year={2025},
  school={University of Victoria}
}

@inproceedings{Cook_2020,
  title={The Quantum Alternating Operator Ansatz on maximum $k$-vertex cover},
  author={Cook, Jeremy and Eidenbenz, Stephan and B{\"a}rtschi, Andreas},
  booktitle={2020 IEEE International Conference on Quantum Computing and Engineering (QCE)},
  pages={83--92},
  year={2020}
}

@inproceedings{Bartschi_2020,
  title={Grover mixers for {QAOA}: Shifting complexity from mixer design to state preparation},
  author={B{\"a}rtschi, Andreas and Eidenbenz, Stephan},
  booktitle={2020 IEEE International Conference on Quantum Computing and Engineering (QCE)},
  pages={72--82},
  year={2020}
}

@inproceedings{chalupnik2022augmenting,
  title={Augmenting {QAOA} Ansatz with Multiparameter Problem-Independent Layer},
  author={Chalupnik, Michelle and Melo, Hans and Alexeev, Yuri and Galda, Alexey},
  booktitle={2022 IEEE International Conference on Quantum Computing and Engineering (QCE)},
  pages={97--103},
  year={2022}
}

@article{chandarana2022digitized,
  title={Digitized-counterdiabatic quantum approximate optimization algorithm},
  author={Chandarana, Pranav and Hegade, Narendra N and Paul, Koushik and Albarr{\'a}n-Arriagada, Francisco and Solano, Enrique and Del Campo, Adolfo and Chen, Xi},
  journal={Physical Review Research},
  volume={4},
  number={1},
  pages={013141},
  year={2022},
  publisher={APS}
}

@article{pelofske2023solvingclique,
  title={Solving larger maximum clique problems using parallel quantum annealing},
  author={Pelofske, Elijah and Hahn, Georg and Djidjev, Hristo N},
  journal={Quantum Information Processing},
  volume={22},
  number={5},
  pages={219},
  year={2023},
  publisher={Springer}
}

@article{skolik2021layerwise,
  title={Layerwise learning for quantum neural networks},
  author={Skolik, Andrea and McClean, Jarrod R and Mohseni, Masoud and Van Der Smagt, Patrick and Leib, Martin},
  journal={Quantum Machine Intelligence},
  volume={3},
  number={1},
  pages={5},
  year={2021},
  publisher={Springer}
}

@article{zhu2025optimizing,
  title={Optimizing Minimum Vertex Cover Solving via a GCN-assisted Heuristic Algorithm},
  author={Zhu, Enqiang and Bao, Qiqi and Zhang, Yu and Liu, Chanjuan},
  journal={arXiv preprint arXiv:2503.06396},
  year={2025}
}

@inproceedings{gao2016fixed,
  title={Fixed-parameter single objective search heuristics for minimum vertex cover},
  author={Gao, Wanru and Friedrich, Tobias and Neumann, Frank},
  booktitle={International Conference on Parallel Problem Solving from Nature},
  pages={740--750},
  year={2016},
  organization={Springer}
}

@article{cai2017finding,
  title={Finding a small vertex cover in massive sparse graphs: Construct, local search, and preprocess},
  author={Cai, Shaowei and Lin, Jinkun and Luo, Chuan},
  journal={Journal of Artificial Intelligence Research},
  volume={59},
  pages={463--494},
  year={2017}
}

\end{document}